\documentclass[10pt,twocolumn]{article} 
\usepackage{graphicx}
\usepackage{amssymb}
\usepackage{url,hyperref}
\usepackage{pgfplots}
\usepackage{verbatim}
\usepackage{amsmath}
 \usepackage{multirow}

\begin{document}

\title{Auto-detecting groups based on textual similarity for group recommendations}

\author{Chintoo Kumar, C. Ravindranath Chowdary \\
\\
Department of Computer Science and Engineering \\
Indian Institute of Technology (BHU), Varanasi, India, 221005 \\
\\
\\
chintookr.rs.cse17@itbhu.ac.in, rchowdray.cse@iitbhu.ac.in  \\
}

\maketitle
\thispagestyle{empty}

\begin{abstract}
In general, recommender systems are designed to provide personalized items to a user. But in few cases, items are recommended for a group, and the challenge is to aggregate the individual user preferences to infer the recommendation to a group. It is also important to consider the similarity of characteristics among the members of a group to generate a better recommendation. Members of an automatically identified group will have similar characteristics, and reaching a consensus with a decision-making process is preferable in this case. It requires users-items and their rating interactions over a utility matrix to auto-detect the groups in group recommendations. We may not overlook other intrinsic information to form a group. The textual information also plays a pivotal role in user clustering. In this paper, we auto-detect the groups based on the textual similarity of the metadata (review texts). We consider the order in user preferences in our models. We have conducted extensive experiments over two real-world datasets to check the efficacy of the proposed models. We have also conducted a competitive comparison with a baseline model to show the improvements in the quality of recommendations. 

\end{abstract}

\section{Introduction}
A recommender system suggests relevant content to a user based on the user's experience, content, and interactions between the user and items \cite{Ricci2011,aggarwal2016recommender,Aytekin2019,tran2020recommender,MFrecsys20}.
There are some scenarios where more than a person with diversified interests are involved in a recommendation process. The objective of a group recommendation is to provide suggestions that would suit a group of people \cite{AGARWAL2017115,Dara2019,ZAN2021401,KayaBT20}. The composition of a group from a given dataset is important as user groups are usually not predefined. 

Exploring a particular group from the available information is a challenging task. 
The members of the group who share similar characteristics may be unfamiliar with one another. Therefore, we propose to explore a group before generating the recommendation as the group's composition is vital in group recommendations. The formed group is known as an automatically identified group \cite{Boratto2015,BORATTO2017424,YALCIN2021114111,YALCIN2021114709}. The existing models identify and detect the groups automatically by applying traditional clustering approaches, i.e., k-means clustering \cite{Boratto2015,BORATTO2017424,YALCIN2021114111,YALCIN2021114709}, agglomerative clustering \cite{Ntoutsi2012} and Louvain clustering \cite{Blondel2008} over a utility matrix.

A clustering approach over a utility matrix mainly considers attributes: users, items, and ratings to form a group. It does not take account of some other meta-data information to form a group. The review texts are auxiliary information available in most of the datasets. In the proposed work, we consider the textual similarities among review texts to auto-detect a group. In this kind of established group, the generated recommendations are based on the similarity between items, and therefore, the overall group satisfaction score is higher. 

The key contributions of our work are as follows:
\begin{itemize}
\item To the best of my knowledge, this is the first attempt to reuse a pre-trained language model to auto-detect the groups in group recommendations.
\item We analyzed the obtained automatically identified groups using a cluster validation approach.
\item We study the effectiveness of the proposed models based on the order in user preferences. We also conducted a comparative study with a baseline model.  
\end{itemize}

\section{Related work}
\label{sec:previous work}

In \cite{Ntoutsi2012}, authors proposed the model to recommend items to a group based on the past users' experience, i.e., similar users who liked items in the past are likely to give the same preferences in the present. In \cite{BasuRoy2015}, the authors composed a group using the proposed consensus function that comprises the importance of group relevance and group disagreement factors equally. Authors in \cite{Ortega2016} first incorporate a matrix factorization-based model in a group recommender system to map the users and items into a latent space and generate the recommendations using collaborative filtering on various group sizes. 
In \cite{Boratto2015,BORATTO2017424}, authors proposed a set of group recommender systems by using the K-means 
clustering algorithm that automatically detects groups of users. The introduced \textit{Predict \& Cluster} model firstly predicts the unknown rating using collaborative filtering, and the K-means algorithm is employed over the utility matrix to form a group.

In \cite{AGARWAL2017115}, we introduced the concept of the ordering of items in group recommender systems. We introduced the \textit{User Satisfaction with order (USO)} function to justify the importance of order in the group recommendation. The variants of aggregated voting \cite{AGARWAL2017115,Masthoff2011} and least misery \cite{AGARWAL2017115,Masthoff2011} methods are proposed to maximize the overall group satisfaction with the generated recommendation vector. The group is formed based on overlap similarity scores. Since we need to compute correlations among all users when identifying the group of users, the group formation technique is computationally inefficient and not preferable. In \cite{Dara2019A}, we introduced the model based on flexibility in user preferences. We adopted the similar consensus functions of ordered preference-based model \cite{AGARWAL2017115} to establish this concept.

In \cite{YALCIN2021114111}, authors introduced a new aggregation technique based on the concept of entropy and used \textit{K-means} to detect an automatically identified group. Further, authors in \cite{YALCIN2021114709} used the bisecting K-means algorithm for automatic group identification. The framework introduced in \cite{groupIM20} can make item recommendations to an ephemeral group efficiently. Authors in \cite{BARZEGARNOZARI2020106296} composed groups using fuzzy-c means (FCM) clustering, and the users are potential members of the formed group based on the pearson correlation coefficient measure. 

Although existing literature presents the group formation using traditional clustering approaches over the utility matrix, none of the methods considers semantic similarity on the textual information to form the groups. Therefore, the proposed group formation technique is a promising approach to improving the quality of group recommendations.

\section{Proposed work}
\label{sec:methodology}
Most of the existing group recommender systems focus on improving the quality of recommendation by learning the representation of users and items and modelling their interaction \cite{99,100,groupIM20}. But, these approaches do not pay much attention to compare users in a group. It is more advantageous to compose a group before making the recommendation in order based user preference models. Users who gave the same rating to the same items in a utility matrix become members of an automatically identified group. We assume that users who have semantically similar reviews on the same products could be used for auto-detecting a group. Each group member would play an equal role in the decision-making process and improve the overall group satisfaction score.

Most of the group recommender systems initial step is to partition members into groups automatically. This particular classification of the group is known as an automatically identified group \cite{BORATTO2017424,Boratto2015,YALCIN2021114111}. Our focus is to detect these kinds of groups automatically from the available resources. In a utility matrix, a user has to rate an item.  Till now,  the user-item and rating interactions over a  utility matrix are used for automatic group identification. However, while forming a group, they overlook other intrinsic information of a dataset. We auto-detect groups based on the metadata (review texts given by users on items) in this work.  The textual similarities among the users provide a potential group to take part in a recommendation process.
\begin{itemize}
\item \textit{Universal Sentence Encoder}: Transfer learning \cite{5288526} is the reuse of a pre-trained model for a novel task. We can make use of it in a wide range of applications. Of late, there are many pre-trained language models explored to make the natural learning process task easier. \textit{Universal sentence encoder} \cite{cer2018universal} is a pre-trained language model. This model is publicly available in TensorFlow Hub \footnote{https://www.tensorflow.org/hub/tutorials/}. The model is trained on a variety of data sources. It takes an input (English text) of variable length size and produces a 512 dimensional vector as an output. 
\end{itemize}
We use this pre-trained language model for sentence embedding. We embed the review texts using the universal sentence encoder. We get a matrix of dimension $m\times n$ where $m$ is the total number of sentences (review texts), and $n$ is a vector of size 512 after embedding.
 
Semantically similar sentences will have similar weight scores after embedding. We calculate the similarity scores among the embedded sentences vectors using cosine similarity. Then, we get a similarity matrix. Finally, we apply spectral clustering \footnote{https://scikit-learn.org/stable/modules/generated/sklearn.cluster.SpectralClustering.html} approach on the similarity matrix to auto-detect a cluster for the instances. If a user has provided reviews on more than one product in the original dataset, then the corresponding cluster varies. We took the majority of the labeled clusters for a particular user to assign it.  

We perform cluster validation over the obtained results. The most popular index is the Adjusted Rand Index (ARI) \footnote{//scikit-learn.org/stable/modules/generated/sklearn.metrics.adjusted\_rand\_score.html} \cite{CAMPELLO2007833,TALBURT201163} for cluster analysis. The ARI can get a value ranging from -1 to 1. The higher the value, the better it matches the original data.
\begin{equation}
 ARI=\frac{RI-Expected (RI)}{Max(RI)-Expected(RI)}
\end{equation}
Where \textit{RI} is the rand index\footnote{https://scikit-learn.org/stable/modules/generated/sklearn.metrics.rand\_score.html}. The rank index \cite{rand1971objective,CAMPELLO2007833} is a metric to evaluate the quality of clustering result. The range of the rand index is [0,1]. The value `0' signifies that the two data clusters do not agree on any pair of points, while the value `1' signifies that both the clusters are the same. The Rand Index (RI) is defined as follows:
\begin{equation}
RI=\frac{TP+TN}{TP+FP+FN+TN}=\frac{a+d}{\binom{n}{2}}
\end{equation}
Where
\begin{enumerate}
 \item TP (True Positive) : A true positive decision assigns two similar documents to the same cluster.
 \item TN (True Negative): A true negative decision assigns two dissimilar documents to different clusters.
 \item FP (False Positive): A false positive decision assigns two dissimilar documents to the same cluster.
 \item FN (False Negative): A false negative assigns two similar documents to different clusters.   
\end{enumerate}

 \begin{figure}
    \centering

	\includegraphics[width=0.8\linewidth]{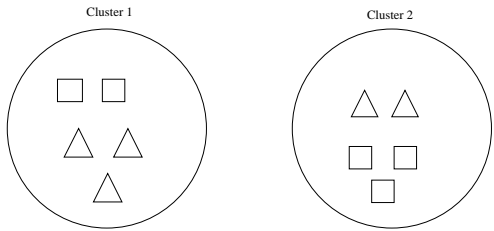}

	\caption{Two clustering assignments}

	\label{fig:ari}

\end{figure}
ARI \cite{CAMPELLO2007833,TALBURT201163} can also be represented as:
\begin{equation}
 ARI=\frac{a-[\frac{(a+c)\times(a+b)}{\binom{n}{2}}]}{\frac{(a+c)\times (a+b)}{2}-[\frac{(a+c)\times(a+b)}{\binom{n}{2}}]}
\end{equation}

Fig. \ref{fig:ari} shows ten data points ($n=10$) into two partitions (cluster 1 and cluster 2). 
Each of the clusters contains 5 data points. First, we compute TP+FP. The total number of positive or pairs of documents that are in the same cluster are:
TP+FP= $\binom{5}{2}+\binom{5}{2}=20$

Out of these, both square and triangle kinds of pairs in both clusters are positive. So,

TP=$\binom{2}{2}+\binom{3}{2}+\binom{2}{2}+\binom{3}{2}=8$; FP=20 - 8=12 and

FN=$2\times 3+ 3\times 2=12$ and TN=$2\times2+3\times3=13$
\begin{table}[]
\caption{Contigency Table}
\label{tab:cont}
\begin{tabular}{|l|l|l|}
\hline
                & Same cluster & Different cluster \\ \hline
Same class      & TP(a)=8         & FN(b)=12             \\ \hline
Different class & FP(c)=12        & TN(d)=13             \\ \hline
\end{tabular}
\end{table}

Table \ref{tab:cont} represents the contingency table.
So, RI=$\frac{21}{45}$=0.46 and ARI=$\frac{8-[\frac{400}{45}]}{20-[\frac{400}{45}]}$=-0.08
While $Precision(P)=\frac{a}{a+c}=\frac{8}{20}=0.4$, $Recall(R)=\frac{a}{a+d}=\frac{8}{21}=0.38$ and $F-measure=\frac{2PR}{P+R}=\frac{2\times 0.4 \times 0.38}{0.4+0.38}=0.39$

Once the community (group) is formed, we need to recommend items that can provide maximum satisfaction to members of the group. We used different consensus functions to generate the recommendation list for a  detected group, which also determine that a user of a group is maximally satisfied with the recommended item-set.
The proposed model recommends items for a group by considering the order in user preferences.
In the ordered preference-based model; first, each member of a group provides preferences in order. Then, the model aggregates individual user preferences using a consensus function to generate a recommendation vector of group budget `k'.

We adopted the variants of least misery and aggregated voting methods in our models to generate the recommendation vectors:
\begin{itemize}
\item {Least Misery Method (LMM)} : We choose the least satisfied user in each iteration, and the item of his/her preference list that maximizes the overall group satisfaction score is added in the recommendation vector.
\item {Least Misery Method with Priority (LMMP)}: The priorities of users are calculated to break a tie to choose a least satisfied user.
\item {GReedy Aggregated Method (GRAM)}: In each iteration, it greedily chooses an item from the user satisfaction score matrix \cite{AGARWAL2017115} that maximizes the overall group satisfaction score and appends that item into the recommendation vector.
\item {Hungarian Aggregated Method (HAM)}: This method is an extension of the aggregated method that provides an optimal solution. Moreover, it produces the best results when we consider a smaller number of users and items. 
\end{itemize}
The size of the preference list would be less than or equal to the group budget in the flexible user preference-based model \cite{Dara2019A}. Users of a group provide at least one item or a maximum of $k$ items in the preference lists, and finally, the model generates a recommendation vector of size $k$. We adopt the similar consensus functions of our previous work \cite{AGARWAL2017115} in the proposed models. 
\section{Experimental setup and result analysis}
\label{sec:result}
We took two real-world datasets: Amazon-music (Digital Music) \footnote{https://jmcauley.ucsd.edu/data/amazon/} and Clothing Fit Data (Modcloth) \footnote{https://cseweb.ucsd.edu/~jmcauley/datasets.html} for the evaluation of the proposed work. 
\newline\textbf{Amazon-music (Digital Music)}: The digital music dataset holds reviews and metadata information from amazon. The digital music dataset has 5148 distinct reviewers and 2582 distinct asin (a unique number assigned to each product/music). The rating scale (named ``overall'' ) is 1-5. The total number of reviewers and asin (item) interactions over `overall' attribute are 51796 (total instances). The dataset has also review text as metadata. 

\textbf{Clothing Fit Data (Modcloth)}: The dataset contains measurements of clothing fit from ModCloth\footnote{https://modcloth.com/}.
The total number of distinct users are 47958. The total number of distinct products (items) is 1378. The total number of users and items interactions (transactions) are 82790. Each user has rated a product using the attribute ``quality" on the rating scale of 1-5. There is a total of eighteen attributes in the dataset. In addition, the dataset has auxiliary information like ratings and reviews, fit feedback, measurements of users and items, category, etc.

\subsection{Pre-processing step:}
In the Amazon-music and Clothing Fit Data (Modcloth) datasets, we dropped irrelevant attributes (attributes that do not feed as inputs in the proposed model) from both datasets, respectively. It does not play any role in our model during the data preprocessing task. We consider the order of items/asin as they appear in the datasets.  
In the case of Amazon-music dataset, there is negligible empty review text from a particular user. So, we replaced NaN entry of the attribute `reviewText' with a stopword.
But, in the case of the Modcloth dataset, many entries in the attribute `review\_text' are empty, i.e., assigned with NaN in instances. So, we dropped those instances. So, finally, the total number of users and items are 44856 and 1324, over 76065 transactions.

\subsection{Group formation step:} 
We use the spectral clustering method over the obtained affinity matrix after getting the embedded sentence vectors using the universal sentence encoder and applied cosine similarity to calculate the similarity among review texts. The given datasets have text reviews in the English language. So, we used the universal sentence encoder for sentence embedding. In the case of multilingual text, we can use the multilingual universal sentence encoder \cite{muse19}. We took 500 users in each dataset to partition into clusters. Then, we applied silhouette coefficient \cite{ROUSSEEUW198753} to get the optimal number of clusters.
Silhouette coefficient ($s$) can be defined as:
\begin{equation}
s=\frac{y-x}{max(x,y)}
\end{equation}
Where $x$ is the mean distance between a sample and all other points in the cluster.

$y$ is the mean distance between a sample and all other points in the next nearest cluster. 
We got the number of optimal clusters as 2 in both datasets. 

To measure the effectiveness of formed groups, we compare them to the baseline method. In baseline, we use \textit{Predict \& Cluster} (P\& C) model \cite{BORATTO2017424,BorattoArt2015} to check the efficacy of our proposed model. At baseline, firstly, we calculate the scores of unrated items using collaborative filtering \cite{Schafer2007,Salakhutdinov_rbm2007} then applied the K-means technique for group formation. The hyper-parameters in the K-means-based model are the number of clusters ($k=2$), seeds(initial value=0), and euclidean distance as distance measure. 
\subsection{Recommendation step:}  
Score vector holds values based on user satisfaction with order \cite{AGARWAL2017115} function.
To generate the score vector \cite{AGARWAL2017115}, we took values of $a$  and regularization factor $c$ as 2 and 1 respectively. It determines the score based on the index by taking the absolute difference between the position of the item $i$ of the user preference vector and the recommendation vector's item position $p$. The lower the difference, the higher the value it contains. 
  
It has also been observed that the number of user preferences of a group in the Clothing Fit Data dataset is lesser than that of the Amazon-music dataset. In the proposed model, the preferences of each member of the group are in order. Then, the model aggregates individual user preferences using a consensus function. Finally, the generated recommendation vector holds items of group budget `k'  (where `k' is the maximum number of items that can be recommended ) based on the proposed evaluation metric: group satisfaction score. We used aggregation strategies, e.g., LMM, LMMP, GRAM and HAM to produce the recommendation for an automatically identified group. As a result, the proposed models maximize the overall group satisfaction score compared to the baseline method.

\subsection{Results and discussions}
To find the effectiveness of automatically detected groups using the proposed approach, we compare the result with the baseline model. We use an external metric for cluster validation. In the Amazon-music dataset, the rand index (RI) and adjusted rand index (ARI) scores are 0.5 and 0.046, respectively. The rand index score is much higher than the adjusted rand index score and validates the obtained clusters perfectly. We used the adopted versions of two widely known versions of consensus functions for recommendation: aggregated voting (AV) and least misery to prove the efficiency of the proposed system. The adopted versions that we used in this paper are least misery method (LMM), least misery method with priority (LMMP), greedy aggregated method (GRAM) and the hungarian aggregated method (HAM). 
 \begin{figure}
    \centering

	\includegraphics[width=\linewidth]{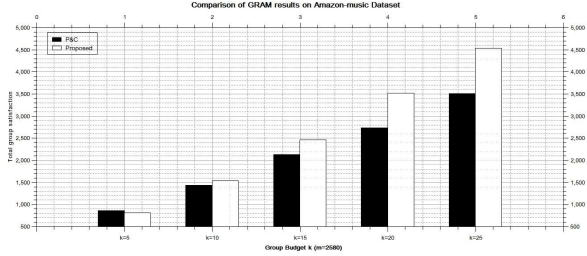}

	\caption{GRAM results on Amazon-music dataset}

	\label{fig:GRAMAM}

\end{figure}

 \begin{figure}
    \centering

	\includegraphics[width=0.8\linewidth]{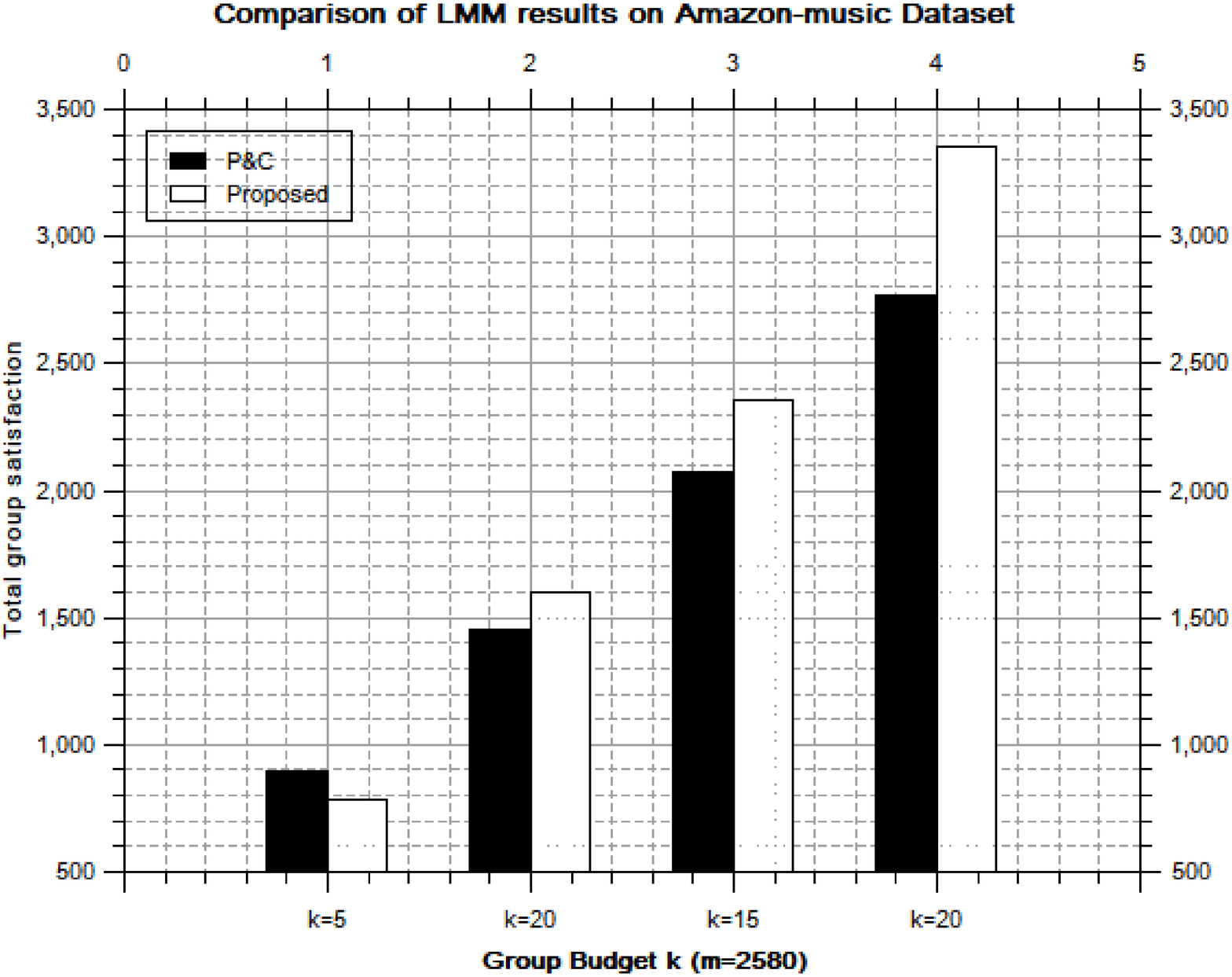}

	\caption{LMM results on Amazon-music dataset}

	\label{fig:LMMAM}

\end{figure}
 \begin{figure}
    \centering

	\includegraphics[width=\linewidth]{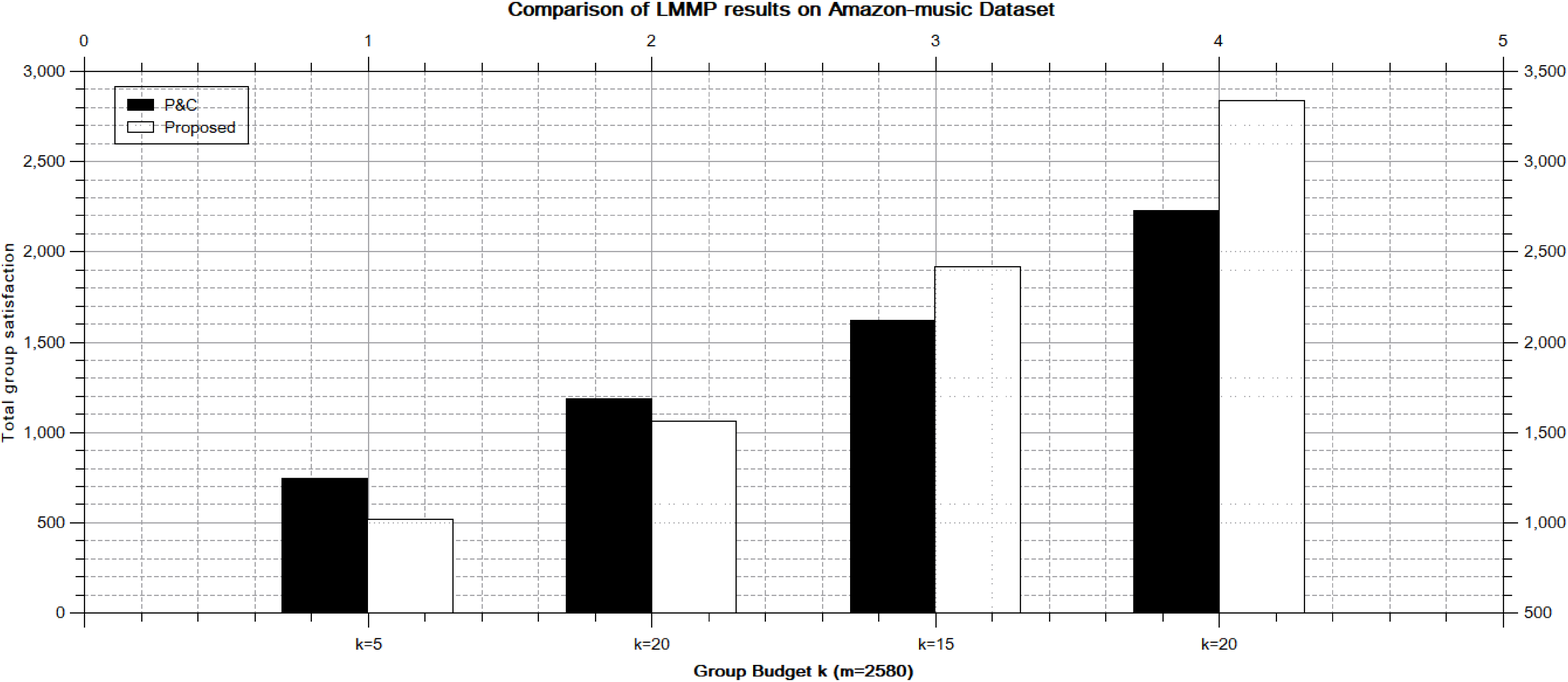}

	\caption{LMMP results on Amazon-music dataset}

	\label{fig:LMMPAM}

\end{figure}

\begin{figure}
    \centering

	\includegraphics[width=0.8\linewidth]{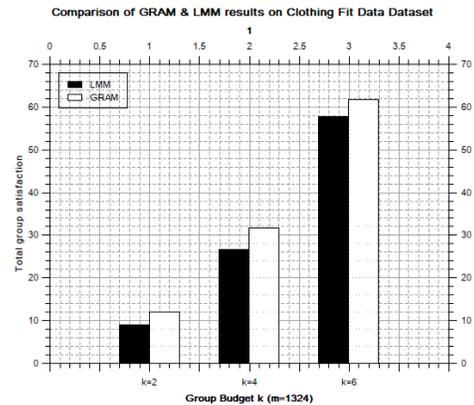}

	\caption{ GRAM \& LMM results on Clothing Fit Data Dataset}

	\label{fig:modcloth}

\end{figure}

\begin{figure}
    \centering

	\includegraphics[width=\linewidth]{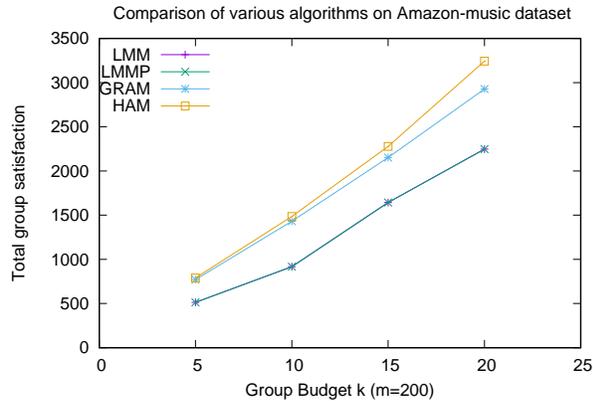}

	\caption{Results using various consensus functions on Amazon-music Dataset}

	\label{fig:AlgosAM}

\end{figure}
We applied different consensus functions in automatically detected groups and measured the performance by varying values of group budget $(k)$ and number of items $(m)$.  We conducted a comparative study between a baseline model to check the overall group satisfaction score improvement. The consensus functions generated group scores, which reflect the interests and preferences of all the group members. Figure \ref{fig:GRAMAM}, \ref{fig:LMMAM} and \ref{fig:LMMPAM} show that the proposed model outperform compare to baseline model.

There is a positive impact on the results of the  proposed model by varying group budget (k) and the number of items (m). It is evident from our experiments that a higher value of $k$  provides a greater satisfaction. Fig. \ref{fig:GRAMAM}, Fig. \ref{fig:LMMAM}, Fig. \ref{fig:LMMPAM}, Fig. \ref{fig:modcloth} and Fig. \ref{fig:AlgosAM} show this characteristic.
The overall group satisfaction increases with the increase in the number of items $(m)$.  Fig. \ref{fig:GRAMAM} and Fig. \ref{fig:AlgosAM} validate this characteristic. We can conclude that consensus functions like LMM, LMMP, GRAM and HAM on the proposed model results in better group satisfaction by considering ordering in user preferences. The proposed group formation technique has a positive role in improving the overall group satisfaction score. 
\section{Conclusions}
\label{sec:Conclusion}
In this paper, we auto-detect the group using metadata like text reviews and then provide recommendations for users of the automatically identified group. Furthermore, it maximizes the overall group satisfaction score when we auto-detect the group based on textual similarity and consider the order in user preference based models.
The extensive experiments over datasets show that review texts are auxiliary information for user clustering. A system needs to deal with scalability and sparsity problem over a utility matrix to compose a potential group but using this approach to form a group is more desirable. It may alleviate the sparsity problem to a great extent. The proposed method is a promising approach to improve the quality of group recommendations. 
\newpage


\end{document}